\documentclass[slac_one]{revtex4}
\usepackage{graphicx}
\usepackage{fancyhdr}
\newcommand{\met}{\,/\!\!\!\!E_{T}}

\newcommand{\sht}{H_T}

\newcommand{\tprime}{t^\prime}
\newcommand{\tprimebar}{\bar{t}^\prime}

\begin{document}

\title{Search for Heavy Top-like Quarks \boldmath{$\tprime\to Wq$} Using Lepton Plus Jets Events in
      1.96-TeV \boldmath{$p\bar{p}$} Collisions} 

\author{A. Lister (on behalf of the CDF collaboration)}
\affiliation{University of California at Davis, CA 95616, USA}

\begin{abstract}
We present the results of a search for a new heavy top-like quark,
$\tprime$, decaying to a $W$ boson and another quark using the CDF~II
Detector in Run~II of the Tevatron $p\bar{p}$ collider.  New top-like
quarks are predicted in a number of models of new physics.  Using a
data sample corresponding to 2.8~fb$^{-1}$ of integrated luminosity we
fit the observed spectrum of total transverse energy and reconstructed
quark mass to a combination of background plus signal.  We see no
evidence for $\tprime$ production, so use this result to set limits
on the $\tprime\tprimebar$ production cross section times the
branching ratio of $\tprime$to $Wq$ and infer a lower limit of 311
GeV/$c^2$ on the mass of the $\tprime$ at 95\% CL.
\end{abstract}

\maketitle

\thispagestyle{fancy}


\section{INTRODUCTION}

We present the result of a search for a new heavy top-like
quark, $\tprime$, assumed to decay to $Wq$.  Such would be the
case with a ``sequential'' fourth generation in which the bottom-like
quark, $b^\prime$, has mass such that $m(b^\prime)+m(W) > m(\tprime)$.
A fourth chiral generation of massive fermions with the 
same quantum numbers as the known fermions is predicted in a number of models. It is favored by flavor 
democracy~\cite{flavor_democracy}, for example, and arises by unifying spins 
and charges in the GUT SO(1,13) framework~\cite{GUT}. 
Many theoretical models, some of which are discussed in the $\tprime$ search using 0.76 $fb^{-1}$ of data~\cite{tprime_prl}, predict the presence of new heavy quarks that decay predominantly to $Wq$. For this reason, we keep the analysis scope broad and do not focus on a specific model.

The mass limits on $\tprime$ production quoted assume 100\% branching ratio to $Wq$.

\section{ANALYSIS METHOD}
We search for strong pair production of a new
heavy $\tprime$ quark with its associated antiquark, each decaying to
$Wq$, using the large data set in CDF from Run II of the Tevatron.
As in the case of $t\bar{t}$ production, this would lead to events
with leptons, jets, and missing transverse energy.  Employing a
technique based on event kinematics avoids imposing a $b$-quark
tagging requirement, which would limit us to the decay mode
$\tprime\to Wb$.  We select events with a lepton ($e$ or $\mu$),
missing transverse energy, and four or more hadronic jets.  In the
standard model these events topologies can arise from $t\bar{t}$
production, $W$ plus jets production, and multi-hadronic-jet
production from QCD processes.  The observed distribution of total
transverse energy, $\sht$, and reconstructed $\tprime$ mass, $M_{rec}$, allows
discrimination of the $\tprime$ signal from these backgrounds.

We perform a binned likelihood fit of background and signal to the
observed two-dimentsional distribution of $\sht$ and $M_{rec}$.  In
this plane, the $\tprime\tprimebar$ events tend to have larger $\sht$
and $M_{rec}$ than the backgrounds, especially as the $\tprime$ mass
gets larger.  We calculate the likelihood as a function of $\tprime$
signal cross section times branching ratio ($\sigma\cdot B$) and apply
Bayes' Theorem assuming a uniform prior in $\sigma\cdot B$ to obtain a
95\% CL limit on the rate.

\section{EVENT SELECTION}
Selected lepton plus jets events must have an $e$ or $\mu$ having reconstructed
$p_T$ above 20 GeV/$c$, four or more jets with $E_T$ exceeding 15 GeV,
having $|\eta|<2$~\cite{def-eta}, and missing transverse energy $\met > 20$ GeV. To
insure that leptons and jets are reconstructed from the same
interaction, the event $z$ vertex is required to be within 5 cm of that
of the lepton. In addition, we require that the muons should not be back-to-back with the missing transverse energy: $\Delta \phi(\mu \rightarrow \met) \leq 3.05$; this cut removes possibly mis-measured muons. A cut on the leading jet $E_T$ is applied at 60 GeV which removes a large fraction of the QCD and W plus jets backgrounds yet keeps over 95\% of the signal.   

For each event we calculate the mass $M_{rec}$ of the $\tprime$ and
$\tprimebar$ using the same approach as in the measurement of the top
quark mass~\cite{topmass}. Of the possible combinations, we select the
one with the lowest $\chi^2$ for the hypothesis $\tprime\to Wq'$,
demanding that the transverse momenta of the $\tprime$ and
$\tprimebar$ balance, the reconstructed $\tprime$ and $\tprimebar$
masses are equal, and the $W$ mass hypothesis is satisfied by the
relevant jet pair on one side and by the lepton and $\met$ on the
other.


We use observed data to estimate the QCD
contribution, following the same method as in the top cross section
measurement.~\cite{top-kin} We use the \textsc{alpgen}~\cite{alpgen}
Monte Carlo generator to simulate $W$ plus jets events, and the
\textsc{pythia} event generator to simulate both $t\bar{t}$ and
$\tprime\tprimebar$ events.


\section{SYSTEMATIC UNCERTAINTIES}
Imperfect knowledge of various experimental parameters leads to
systematic uncertainties which degrade our sensitivity to a $\tprime$
signal.  All systematic uncertainties are represented by ``nuisance''
parameters in the likelihood, and at each point in $\sigma\cdot B$ we
maximize the likelihood with respect to the values of the nuisance
parameters, most of which are gaussian-constrained to particular
values.

The largest systematic uncertainty is that due to imperfect knowledge
of the jet energy scale.  The nuisance parameter representing this
effect controls how the $\sht$-$M_{rec}$ distribution changes
(``morphs'') as the jet energy scale changes within its uncertainties.
This morphing includes both shape and normalization uncertainties
simultaneously.

Another important systematic uncertainty is due to the lack of
knowledge of the appropriate $Q^2$ scale at which the $W$ plus jets
processes should be evaluated.  We take the larger of the shifts in
apparent $\sigma\cdot B$ due to changing the $Q^2$ scale from the
nominal choice of $p_T$ by a factor of two as the magnitude of this
uncertainty.

Other systematic effects include those due to inperfect knowledge of
the integrated luminosity (5.9\%), the lepton identification
efficiencies (0.7\%), and the QCD background normalization (50\%).
As we have poor knowledge of the $W$ plus jets cross section,
this parameter is free to float in the binned likleihood fit.

\section{RESULTS}
\begin{figure}
  \begin{center}
     \includegraphics[width=88mm]{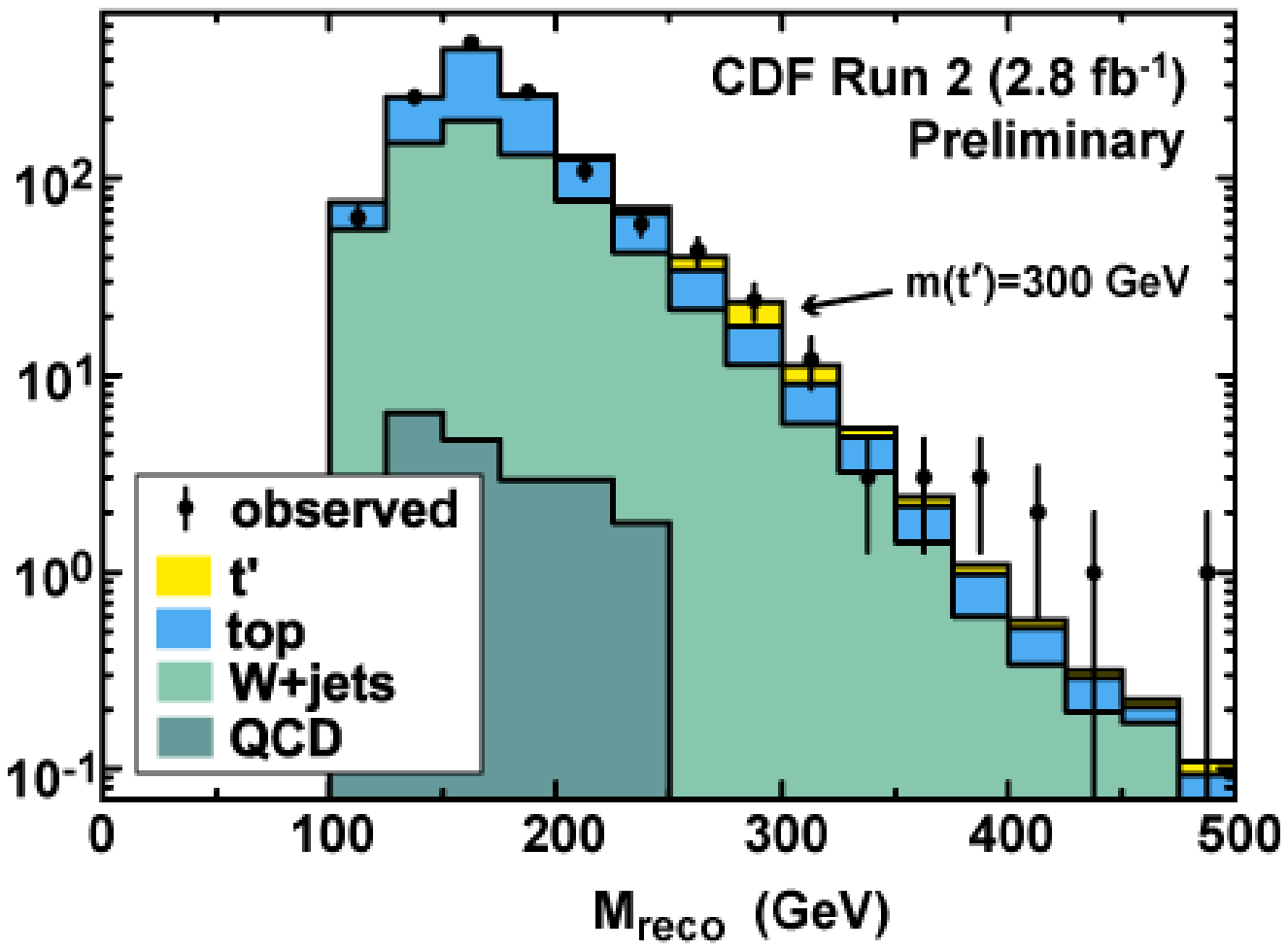}
          \includegraphics[width=88mm]{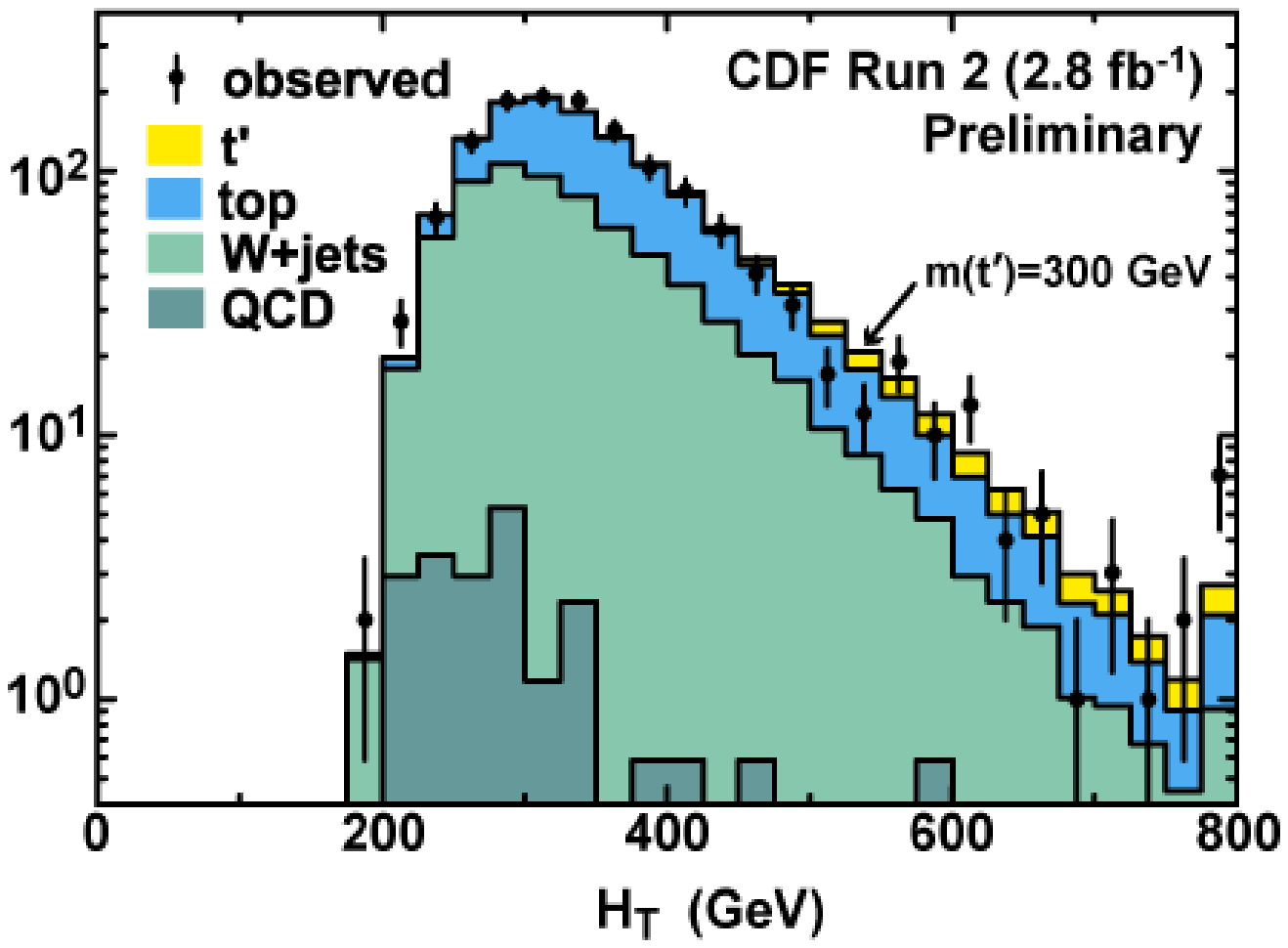}
  \end{center}
  \caption{Observed and predicted distribution of $M_{rec}$ (left) and $\sht$ (right). The predicted
           distribution corresponds to that for a 300 GeV/$c^2$ mass $\tprime$ 
           signal with a cross section times branching ratio at the 
           95\% CL upper limit.}
  \label{fig-fit}
\end{figure}

The likelihoods as a function of $\sigma\cdot B$ reveal no significant
excess attributable to $\tprime\tprimebar$ production. Figure~\ref{fig-fit} shows the
observed data distributions projected into the $M_{rec}$ and $\sht$
dimensions. The figures compare the observed distributions with the
fit to the background plus a 300 GeV/$c^2$ $\tprime$, where the
$\sigma\cdot B$ of the $\tprime$ signal corresponds to that which we
exclude with 95\% confidence.

\begin{figure}
  \begin{center}
     \includegraphics[width=100mm]{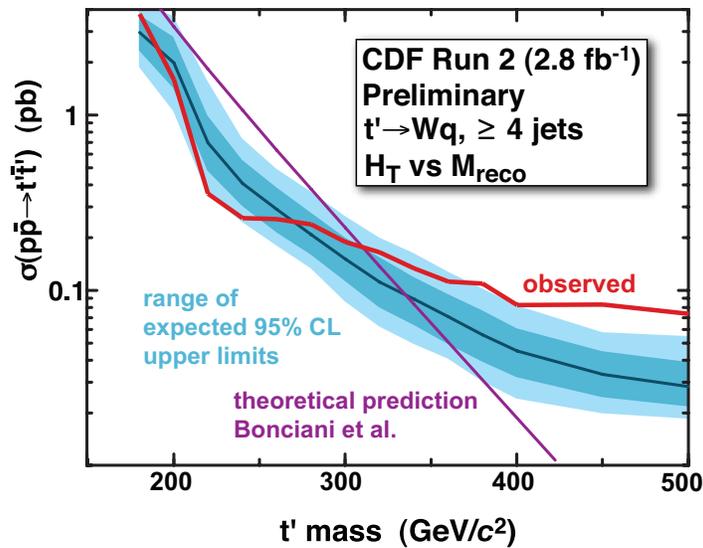}
  \end{center}
  \caption{Observed and expected 95\% CL upper limits on the cross section 
           times branching ratio as a function of $\tprime$ mass. The one and two sigma bands around the expected limits are also shown. The theoretical prediction is shown assuming a 100\%
           branching ratio to $Wq$.}
  \label{fig-limit}
\end{figure}

To obtain a limit on the mass of the $\tprime$ we compare our upper
limit on $\sigma\cdot B$ to the theoretical cross section assuming a
100\% branching ratio $B(\tprime\to Wq)$.  Figure~\ref{fig-limit}
shows these curves, and compres our observed limits to the range of
those expected.  We take the point in $\tprime$ mass where the
observed limit crosses the theoretical cross section as the lower
bound on the mass of the $\tprime$, 311 GeV/$c^2$, at 95\% CL.

So we conclude that the mass of the $\tprime$, if it exists, must exceed 311 GeV/$c^2$ or the $\tprime$
must decay to some other final state.

\begin{thebibliography}{99}
\bibitem{def-eta}     We define the pseudorapidity $\eta$ of a particle
                         three-momentum as $\eta \equiv -\ln(\tan\frac{\theta}{2})$, where
                         $\theta$ is the polar angle of the momentum to the beam direction.
\end{thebibliography}
\end{document}